\documentclass[letter]{aa} 

\usepackage{natbib,booktabs,amsmath,amssymb,mathptmx,mathtools,enumitem,graphicx,hyperref}
\usepackage[flushleft]{threeparttable}
\defcitealias{Bialy2020}{B20}
\usepackage{txfonts}
\hypersetup{
    colorlinks=true,
    linkcolor=blue,
    citecolor=blue,
    filecolor=blue,
    urlcolor=blue
}

\begin{document}

   \title{Constraining the cosmic-ray ionization rate and their spectrum with NIR spectroscopy of dense clouds}
   \subtitle{A test-bed for JWST}
 
   \author{Shmuel Bialy
          \inst{1},
           Sirio Belli\inst{2},
          \and 
           Marco Padovani\inst{3}
          }

   \institute{Department of Astronomy, University of Maryland, College Park, MD 20742, USA\\
              \email{sbialy@umd.edu}
         \and
             Harvard Smithsonian Center for Astrophysics, 60 Garden st., Cambridge, MA 02138, USA
             \and
             INAF-Osservatorio Astrofisico di Arcetri - Largo E. Fermi, 5 - 50125 Firenze, Italy
             }

   \date{Received: November 8, 2021; Accepted: January 14, 2022}

 
  \abstract
   {Low-energy cosmic-rays (CRs) control the thermo-chemical state and the coupling between gas and magnetic fields in dense molecular clouds, the sites of star-formation. However, current estimates of the low-energy CR spectrum ($E \lesssim 1$ GeV) and the associated CR ionization rate
are highly uncertain.}
   {We apply, for the first time, a new method for constraining the CR ionization rate and the CR spectral shape using H$_2$ rovibrational lines from cold molecular clouds.}
   {Using the MMIRS instrument on the MMT, we obtained deep near-infrared (NIR) spectra in six positions within four dense cores, G150, G157, G163, G198, with column densities $N_{\rm H_2} \approx 10^{22}$ cm$^{-2}$.
}
   {We derive 3$\sigma$ upper limits on the H$_2$ $(1-0)$S(0) line (2.22 $\mu$m) brightness in the range
$I = 5.9 \times 10^{-8}$ to $1.2 \times 10^{-7}$ erg cm$^{-2}$ s$^{-1}$ sr$^{-1}$ for the different targets. 
Using both an analytic model and a numerical model of CR propagation, we convert these into upper limits on the CR ionization rate in the clouds' interior, $\zeta = 1.5$ to $3.6 \times 10^{-16}$ s$^{-1}$, and lower limits on the low-energy spectral slope of interstellar CR protons, $\alpha = -0.97$ to $-0.79$.
We show that while MMT was unable to detect the H$_2$ lines due to high atmospheric noise, 
JWST/NIRSpec will be able to efficiently detect the CR-excited H$_2$ lines, making it the ideal method for constraining the otherwise elusive low-energy CRs, shedding light on the sources and propagation modes of CRs.}
   {}

\titlerunning{Constraining the cosmic-ray ionization rate and their spectrum}
\authorrunning{Bialy, Belli \& Padovani}
   \maketitle
%
   
\section{Introduction}
Low-energy CRs ($E \lesssim 1$ GeV) play an important role in determining the thermochemical and dynamical state of dense molecular clouds and are thus crucially important for star-formation
(see \citealt{Padovani2020} for a recent review).
These CRs penetrate into large cloud depths and provide the main ionization source in the gas. 
This ionization is critical: [1] it is the dominant heating mechanism in the gas \citep{Glassgold2012, Girichidis2020} [2] it introduces coupling of the gas with magnetic fields \citep{Padovani2014, Zhao2020}, and [3] it drives the chemistry resulting in the formation of a rich array of interstellar molecules \citep{Dalgarno2006, Caselli2012, Indriolo2013}.

Despite their importance, the spectrum of low-energy CRs, and the CR ionization rate, hereafter $\zeta$, remain uncertain\footnote{In this paper, $\zeta$ denotes the total (primary+secondary) ionization rate, per H$_2$ molecule.}.
This is because at these energies, direct measurements from Earth and Space are affected by solar modulations  \citep[][see \S \ref{sub: CRIR limit numerical} for an elaborate discussion]{Gloeckler2015, Padovani2018a}.
In the interstellar medium, $\zeta$ has been estimated using spectroscopic observations that measure the abundances of various trace molecules, H$_3^+$, OH$^+$, H$_2$O$^+$, ArH$^+$, etc.,
in combination with detailed chemical models.
These methods yield a range of ionization rates, $\zeta\approx10^{-17}-10^{-15}~{\rm s}^{-1}$ in dense and diffuse Galactic clouds \citep{Guelin1982, VanderTak2000, Indriolo2012, Neufeld2017b, Bialy2019c,  Gaches2019} and as high as $\zeta \approx 10^{-14} - 10^{-12}~{\rm s}^{-1}$
towards Galactic center \citep{LePetit2016}, protostellar cluster \citep{Fontani2017, Favre2018}
and extragalactic sources \citep{Muller2015, Gonzalez-Alfonso2018}.
As these $\zeta$ values are based on observations of rare molecules, they rely on chemical models which in turn introduces uncertainties  (especially in dense clouds), due to uncertainties in the chemical rate coefficients, the limited completeness of the chemical network, and an assumption on the gas volume density.

Recently, \citet[][hereafter \citetalias{Bialy2020}]{Bialy2020} has proposed a new method for deriving $\zeta$ in cold molecular clouds that relies on H$_2$, the main constituent of the gas. 
The idea is that in dense clouds ($N_{\rm H_2} \sim 10^{22}$ cm$^{-2}$) H$_2$ is cold and resides primarily in its ground electronic and rovibrational state. 
Secondary electrons produced by the penetrating CRs excite the rovibrational states of H$_2$, which decay to the ground state through photon emission in the NIR.  
As discussed in \citetalias{Bialy2020}, for the temperatures and densities typical of these clouds ($T \lesssim 30$ K, $n \sim 10^{4}-10^{6}$ cm$^{-3}$), collisional de-excitation is negligible, and thus the flux in the emitted lines is proportional to the CR-H$_2$ excitation rate and also to $\zeta$. 
Thus, H$_2$ rovibrational emission lines  may be used to reliably constrain the ionization rate, without the need of chemical models, or additional assumptions on the gas density and compositions.
\citetalias{Bialy2020} emphasized the importance of four particular lines: $(1-0)$O(2), $(1-0)$Q(2), $(1-0)$S(0), and $(1-0)$O(4) at $\lambda=2.63, 2.41, 2.22, 3.00$ $\mu$m respectively (see their Table 1), for which CR excitation dominates over competing excitation processes.

\begin{figure}
    \centering
    \includegraphics[width=0.45\textwidth]{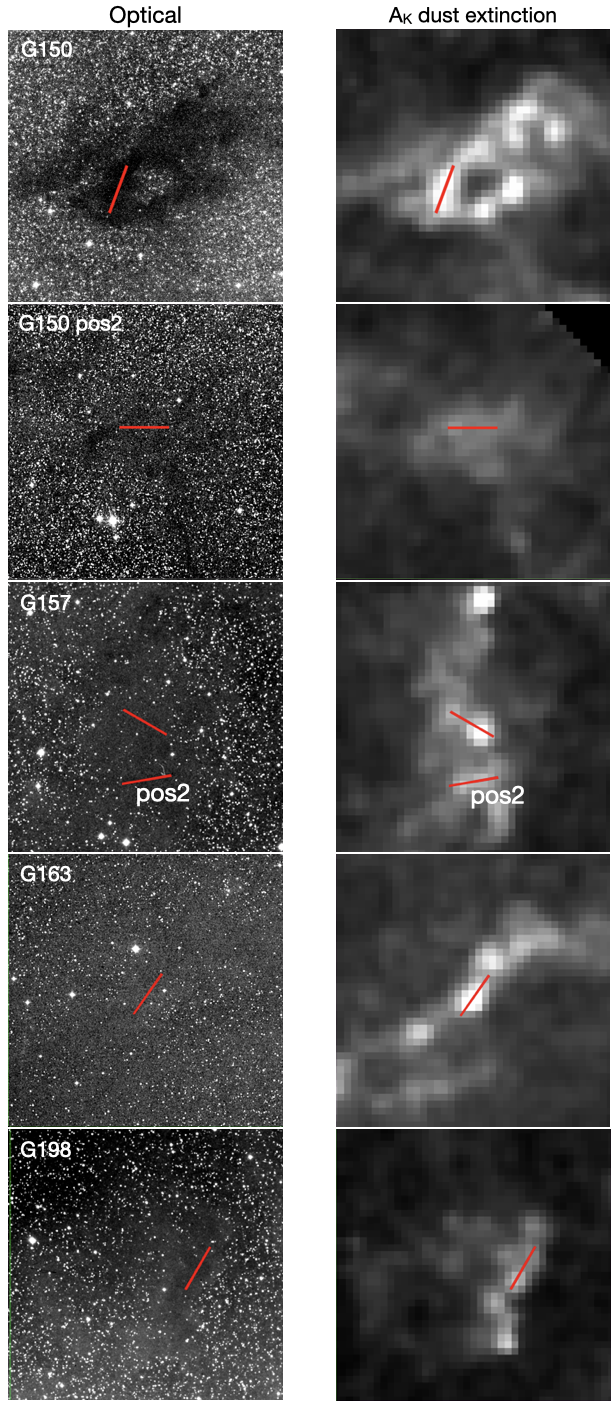}
    \caption{Optical image and NIR dust attenuation map for our target clouds. Images are from ESA-Sky \citep{Baines2017}, based on DSS and 2MASS \citep{Skrutskie2006}. 
    Each panel is $40' \times 40'$. The position of the $7'$ slit is shown in red. 
    }
    \label{fig: targets}
\end{figure}

\begin{figure}
    \centering
    \includegraphics[width=0.4\textwidth]{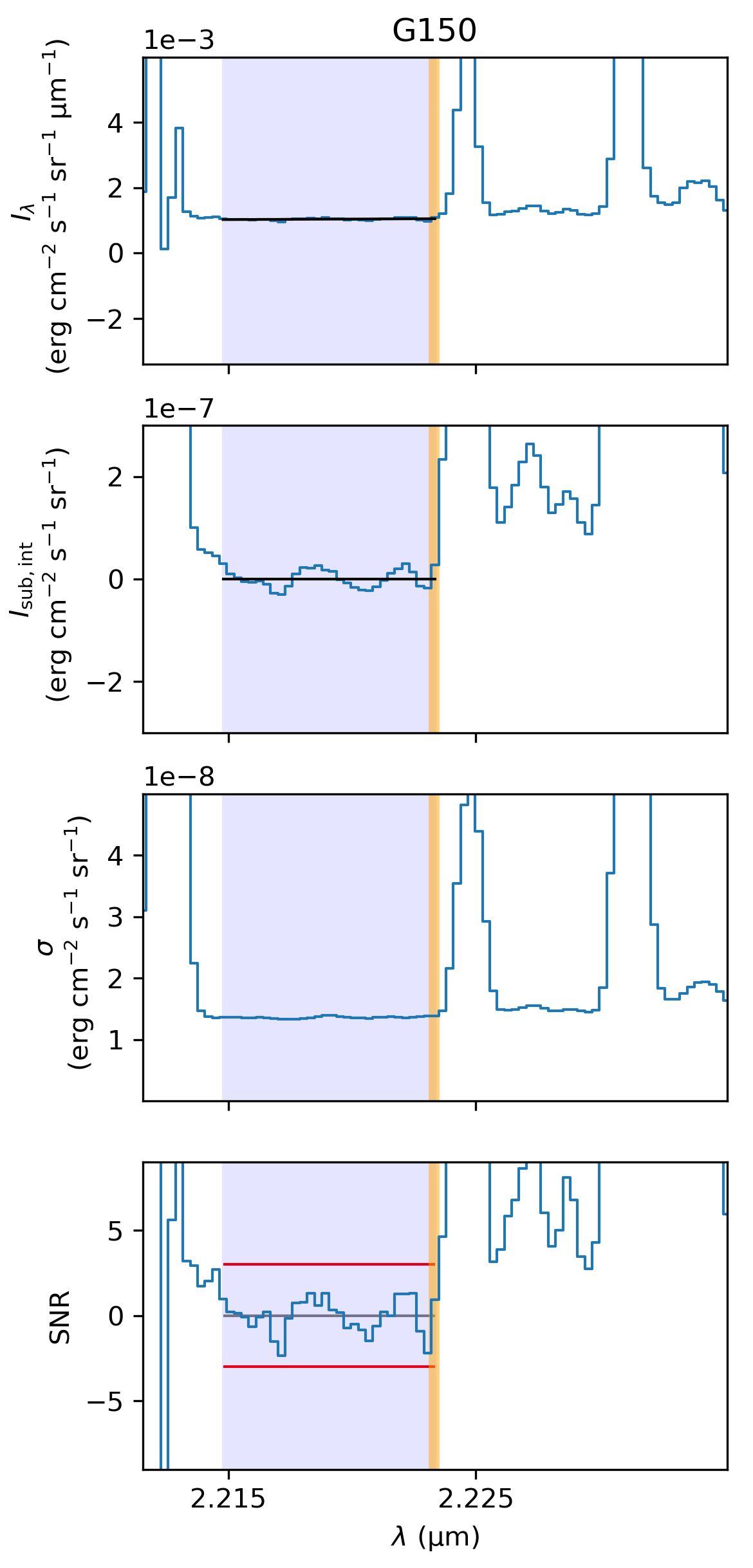}
    \caption{
    Row 1: The 1D spectrum for G150 (see Appendix \ref{app: data reduction} for the rest of the targets).
    The black line is a linear fit to $I_{\lambda}$ in the region with no skylines (blue strip). The vertical orange strip is the wavelength of the $(1-0)$S(0) line.
    Row 2: $I_{\rm sub, int}$, the spectrum after we subtracted from it the linear fit and integrated over the instrument spectral resolution, $\Delta \lambda \approx 7.4 \times 10^{-4} \mu$m.
    Row 3: The error (noise) on $I_{\rm sub, int}$. 
    Row 4: the ${\rm SNR}=I_{\rm sub, int}/\sigma$. Wavelengths are in vacuum.
    }
    \label{fig: spectrum}
\end{figure}

In this paper we report on deep ground-based spectroscopic NIR observations of four nearby dense clouds (\S \ref{sec: observations}).
We use these observations to derive constraints on $\zeta$, and on the interstellar CR proton spectrum at low energies that is impinging on the molecular clouds (\S \ref{sec: CRIR}). 
This is the first time H$_2$ line emission is used to constrain CR properties.
We discuss future prospects for JWST and demonstrate how it will be able to constrain $\zeta$ and the interstellar low-energy CR spectrum and its fluctuations (\S \ref{sec: JWST}).

\section{Observations}
\label{sec: observations}

\subsection{The sample}
\label{sub: targets}

We observed four nearby dense molecular clouds: G150.47+3.93, G157.08-8.69, G163.82-8.44, and G198.58-9.10 (the naming convention is from the Galactic Cold Core, survey; GCC, \citealt{Juvela2012a}). 
Hereafter we use abbreviated names: G150, G157, G163, and G198.
The clouds were selected from the GCC survey based on the following criteria: 
\begin{enumerate}
    \item The cloud should not host, nor be close to, bright stars so that the UV radiation field is weak and the excitation component due to CRs is easier to detect;
    \item The cloud is optically thick in the K band, i.e., $N_{\rm H_2} \gtrsim 10^{22}$ cm$^{-2}$. 
    This maximizes the  emission brightness of the lines. The optical thickness also ensures that the line emission is nearly independent of $N_{\rm H_2}$, as well as reducing contamination from background stars, making the analysis more robust;
    \item The cloud has a large extent on the sky to ensure the entire cloud is covered by the slit, thus maximizing the observed signal;
    \item The cloud has good visibility for a large fraction of the night during the observing period.
\end{enumerate}

\renewcommand{\arraystretch}{2}
\begin{table*}
\centering
  \begin{threeparttable}
    \caption{Observations and Limits on the CR ionization rate and the spectral slope of low-energy CRs}
    \label{table}
		\begin{tabular}{p{10mm} p{19mm} p{10mm} p{15mm} p{13mm} p{22mm} p{18mm} p{18mm} p{16mm}}
			\toprule
			\ \newline \newline \ Cloud Name 
			& \ \newline Coordinates \newline RA-DEC \newline (${\rm J2000}$) 
			&  $d$ \newline distance to cloud \newline (${\rm pc}$) 
			& $N_{\rm H_2}$ \newline H$_2$ column density \newline (${\rm cm}^{-2}$) 
			& $t_{\rm exp}$ \newline Exposure time \newline (${\rm minutes}$) 
			& $I_{\rm (1-0)S(0)}$ \newline $(1-0)$S(0) surface brightness (erg cm$^{-2}$ s$^{-1}$ sr$^{-1}$) 
			& $\zeta$ \newline ionization rate \newline - analytic  \newline (${\rm s}^{-1}$) 
			& $\zeta$ \newline  ionization rate  \newline - numeric  \newline (${\rm s}^{-1}$)
			& $\alpha$ \newline low-energy spectral slope of CR protons \newline 
			\\
			\midrule
			G150 & 04:25:04.0 \newline +54:56:57.1 & 170 & $8.9 \times 10^{21}$  & 211 
			& $\leq 8.3 \times 10^{-8}$  & $\leq 2.3 \times 10^{-16}$ & $\leq 2.1\times 10^{-16}$ & $\geq -0.87$ \\
			G150 (pos.~2) & 04:14:41.8 \newline +55:11:18.3 & 170  & $5.9\times 10^{21}$ & 181 
			& $\leq 1.2 \times 10^{-7}$ & $\leq 4.4 \times 10^{-16}$ & $\leq 3.6 \times 10^{-16}$ & $\geq -0.97$ \\
			G157 & 04:01:39.8 \newline +41:12:20.0 & 450 & $9.9 \times 10^{21}$ & 181 
			& $\leq 6.9 \times 10^{-8}$ & $\leq 1.8\times 10^{-16}$ & $\leq 1.7 \times 10^{-16}$ & $\geq -0.82$ \\
			G157 (pos.~2) & 04:01:38.5 \newline +41:04:03.8 & 450 & $8.0 \times 10^{21}$ & 191 
			& $\leq 5.9 \times 10^{-8}$ & $\leq 1.8\times 10^{-16}$ & $\leq 1.7 \times 10^{-16}$ & $\geq -0.80$ \\
			G163 & 4:25:22.2 \newline +37:09:51.6 & 450 & $1.1 \times 10^{22}$ & 181 
			& $\leq 6.4\times 10^{-8}$ & $\leq 1.5\times 10^{-16}$ & $\leq 1.5 \times 10^{-16}$ & $\geq -0.79$ \\
			G198 & 05:52:18.5 \newline +08:22:45.8 & 445 & $6.8 \times 10^{21}$ & 186 & - & - & - & - \\
			\bottomrule
		\end{tabular}
    \begin{tablenotes}
    \item
      \footnotesize
      (1) The cloud names are abbreviations of the full designations used in the GCC survey \citep{Juvela2012a}: G150.47+3.93, G157.08-8.69, G163.82-8.44, G198.58-9.10, respectively. 
      (2)
      The coordinates correspond to the slit center position.
      (3)
      Distances adopted from \citet[][Table 1]{Juvela2012a}.
      (4)
      H$_2$ columns are based on $A_K$ dust extinction measurements using the NICEST method \citep{lombardi2009}\footnote{\href{http://interstellarclouds.fisica.unimi.it/html/index.html}{http://interstellarclouds.fisica.unimi.it/html/index.html}}, assuming a standard extinction curve \citep{Draine2011}, and averaged along the slit.
      (5)
      Exposure times are the integrated on-source exposures. 
      (6,7,8) The upper limits are based on the 3$\sigma$ noise level at the (1-0)S(0) line wavelength, after applying an additional factor of two multiplicative factor to account for uncertainties in the flux calibration.
      (9) The corresponding lower limits on the low energy spectral slope of interstellar CR protons.
    \end{tablenotes}
  \end{threeparttable}
\end{table*}

\subsection{MMT spectroscopy}
\label{sub:  MMT spectroscopy}
The observations were obtained with the MMIRS instrument on the MMT over several nights between November 2020 and January 2021 using the K3000 grism, with spectral resolution $R\sim3000$.
The G150 and G157 clouds were observed with two different slit positioning, bringing the total number of targets observed to six. The slit placement for each target is shown in Fig.~\ref{fig: targets}, and their properties are listed in Table \ref{table}. 

The upper panel of Fig.~\ref{fig: spectrum} shows a section of the 1D spectrum in the vicinity of the $(1-0)$S(0) line, $\lambda=2.22 \mu$m (denoted by the yellow strip). 
The strong features, including the high peak a few pixels redward of the (1-0)S(0) line, are skylines.
We fit and subtract a linear function (black line) from the spectrum, and further integrate the spectrum over the instrument resolution $\Delta \lambda = \lambda/R \approx 7.4 \times 10^{-4}$ $\mu{\rm m} \approx 2.5$ pixels.
This subtracted-integrated spectrum is shown in Fig.~\ref{fig: spectrum}, 2$^{\rm nd}$ row. 
The 3$^{\rm rd}$ and 4$^{\rm th}$ rows show the noise and the SNR.
At the (1-0)S(0) line wavelength (orange strip), ${\rm SNR }<3$, and we claim a non-detection.
We use 3$\sigma$ (3$^{\rm rd}$ row) noise level at the (1-0)S(0) wavelength to 
place an upper limit on the (1-0)S(0) line brightness for all of our targets.
We have further increased the limits by a factor of two to account for uncertainties in the flux calibration (Table \ref{table}).
See Appendix \ref{app: data reduction} for more details.

 \begin{figure*}
    \centering
    \includegraphics[width=1\textwidth]{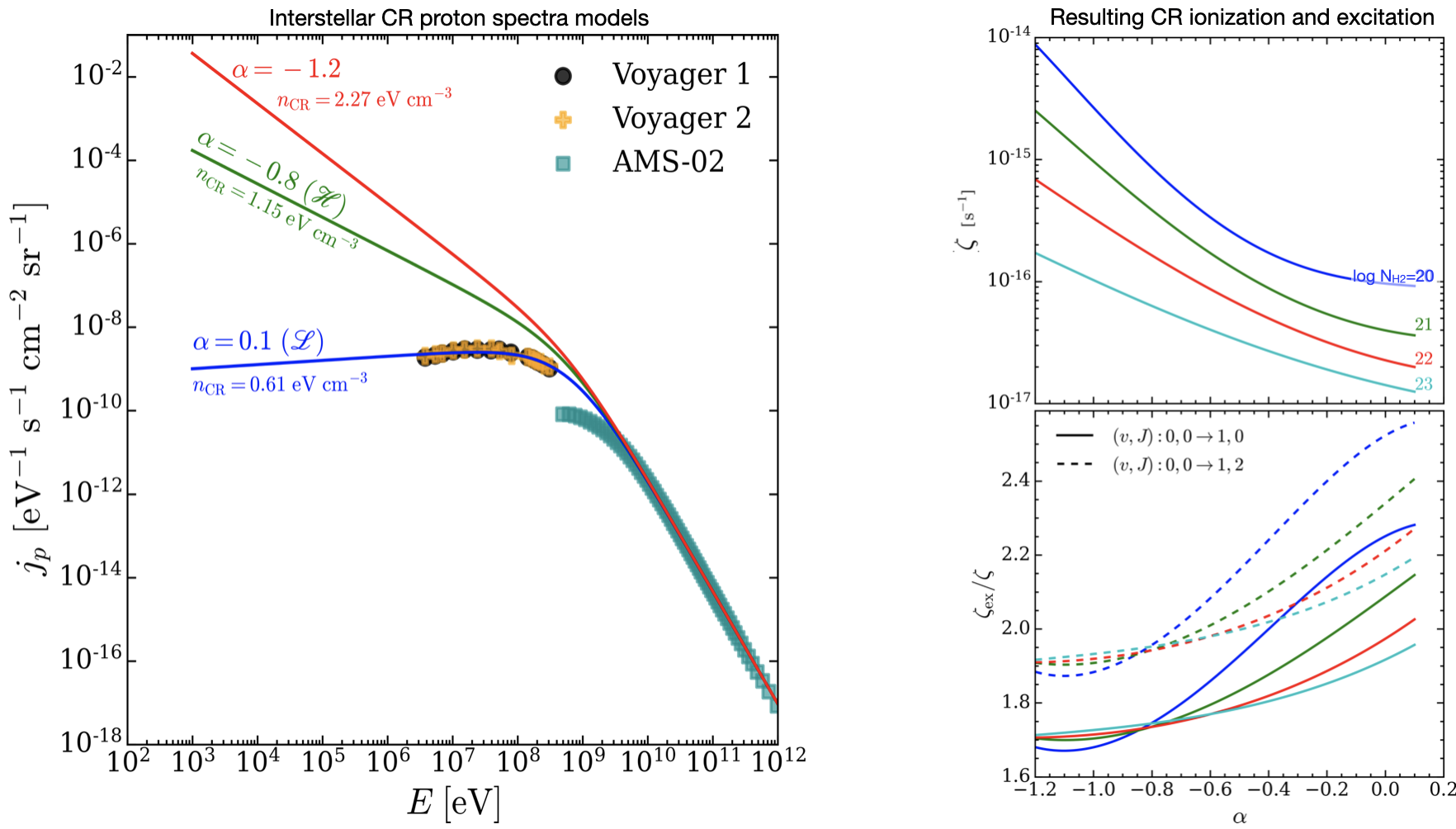}
    \caption{
    Left: the spectrum of interstellar CR protons that are impinging on the cloud at its boundary ($N_{\rm H_2}=0$). For energies $E \gtrsim 1$ GeV, the spectrum is constrained by AMS02 observations \citep{Aguilar2015}. At lower energies, the spectrum is constrain by Voyager observations \citep{Cummings2016, stone2019}, however, since these observations are likely not probing a representative interstellar spectrum (see \S \ref{sub: CRIR limit numerical}), we consider different models with different low-energy spectral slopes, $\alpha$.
    Right: the resulting CR ionization and excitation rates inside the cloud as a function of $\alpha$ and the clouds' molecular column density $N_{\rm H_2}$, as calculated by our CR propagation model.
    }
    \label{fig: ISM spectra}
\end{figure*}

\section{Constraints on the CR ionization rate and CR spectrum}
\label{sec: CRIR}

 In this section we convert our limits on the $(1-0)$S(0) brightness to upper limits on $\zeta$ in the target clouds, and on the spectral slope of low-energy CRs, based on both an analytic and a numerical model.
  
 \subsection{The CR ionization rate - analytic model}
 \label{sub: CRIR limit analytic}
 
 As discussed in \citetalias{Bialy2020} CRs (and secondary electrons produced by CRs) penetrate into molecular clouds and excite the rovibrational levels of  H$_2$ leading to line emission in the NIR. Unlike photo excitation at the cloud surface, or excitation by the H$_2$ formation process, CRs are much more efficient in exciting two specific energy states, the $v=1$, $J=0$ and $v=1$, $J=2$ levels \citep[see also][]{Gredel1995}, resulting in efficient emission of the $(1-0)$S(0), $(1-0)$O(2), $(1-0)$Q(2) and $(1-0)$O(4) lines in the 2-3 $\mu$m range (\citetalias{Bialy2020}).

Since the cloud density is low compared to the levels' critical density, the line emission surface brightness is directly proportional to the CR excitation rate, with
\begin{equation}
\label{eq: I-zeta relation}
  I_{ul} = \frac{1}{4 \pi} gN_{\rm H_2}  \zeta_{\rm ex}  p_{u} \alpha_{(u)l} E_{ul} \ , 
\end{equation}
(Eq.~1 in \citetalias{Bialy2020}, see also Appendix \ref{app: analytic model CR attenuation}).
Here $I_{ul}$ is the line brightness (erg cm$^{-2}$ s$^{-1}$ sr$^{-1}$), $N_{\rm H_2}$ is the H$_2$ column density,
$g \equiv \frac{1-\mathrm{e}^{-0.9N_{22}}}{0.9N_{22}}$ is a factor that accounts for the optical thickness where $N_{22} \equiv N_{\rm H_2}/(10^{22} \ {\rm cm^{-2}})$, and
$\zeta_{\rm ex}$ is the total CR excitation rate, including all H$_2$ levels.
The remaining factors are set by atomic physics: $p_{u}$ is the fraction of all excitations that go onto the specific level of interest, $u$, 
$\alpha_{(u)l}$ is the branching ratio for radiative decay to level $l$ (from upper level $u$), 
and $E_{ul}$ is the energy of the transition.
The subscripts, $u$ and $l$ denote the ``upper" and ``lower" states of the transition.
In our case, for the $(1-0)$S(0) line,  and for CR excitation we have, $u: (v=1, J=2)$, $l: (v=0, J=0)$, $p_{u}=0.47$, $\alpha_{(u)l}=0.3$, $E_{ul}=0.56$ eV (\citetalias{Bialy2020}, Table 1).

The total  H$_2$ CR excitation rate, $\zeta_{\rm ex}$, and the H$_2$ ionization rate, $\zeta$, 
are proportional, with $\varphi \equiv \zeta_{\rm ex}/\zeta \approx 5.8$ \citep{Gredel1995}.
Plugging this back into Eq.~(\ref{eq: I-zeta relation}) and inverting we get: 
\begin{align}
\label{eq: zeta  from I}
 \zeta &=  4 \pi I_{ul} (\varphi p_{u} \alpha_{(u)l} E_{ul} gN_{\rm H_2})^{-1}  \nonumber \\
 &= 2.6 \times 10^{-16} \left(\frac{I_{\rm (1-0)S(0)}}{10^{-7} \ {\rm erg \ cm^{-2} \ s^{-1} \ sr^{-1}}} \right) \frac{0.66}{g} \frac{1}{N_{22}} \ {\rm s^{-1}} \ .
\end{align}
In the numerical evaluation we focused on the $(1-0)$S(0) transition at $N_{22}=1$ (for which $g=0.66$).

For G150,
$N_{22}=0.9$, $g=0.69$, and $I_{\rm S(0)} \leq 8.3 \times 10^{-8}$ erg cm$^{-2}$ s$^{-1}$ sr$^{-1}$ (Table \ref{table}). We obtain a 3$\sigma$ upper limit
$\zeta \leq 2.3 \times 10^{-16}$ s$^{-1}$.
We repeat this exercise for the rest of our observed targets and present the results in Table \ref{table}.

We note that in the original derivation, $\zeta$ does not vary with cloud depth. In practice, as CRs propagate into a cloud they lose energy and $\zeta$ decreases. In the presence of CR attenuation, Eqs.~(\ref{eq: I-zeta relation}-\ref{eq: zeta  from I}) may still be valid under some circumstances (and after applying a correction factor), in which case $\zeta$  represents the attenuated CR ionization rate in the cloud interior. 
See Appendix \ref{app: analytic model CR attenuation}, for more details and limiting cases.


  \begin{figure*}
    \centering
    \includegraphics[width=0.9\textwidth]{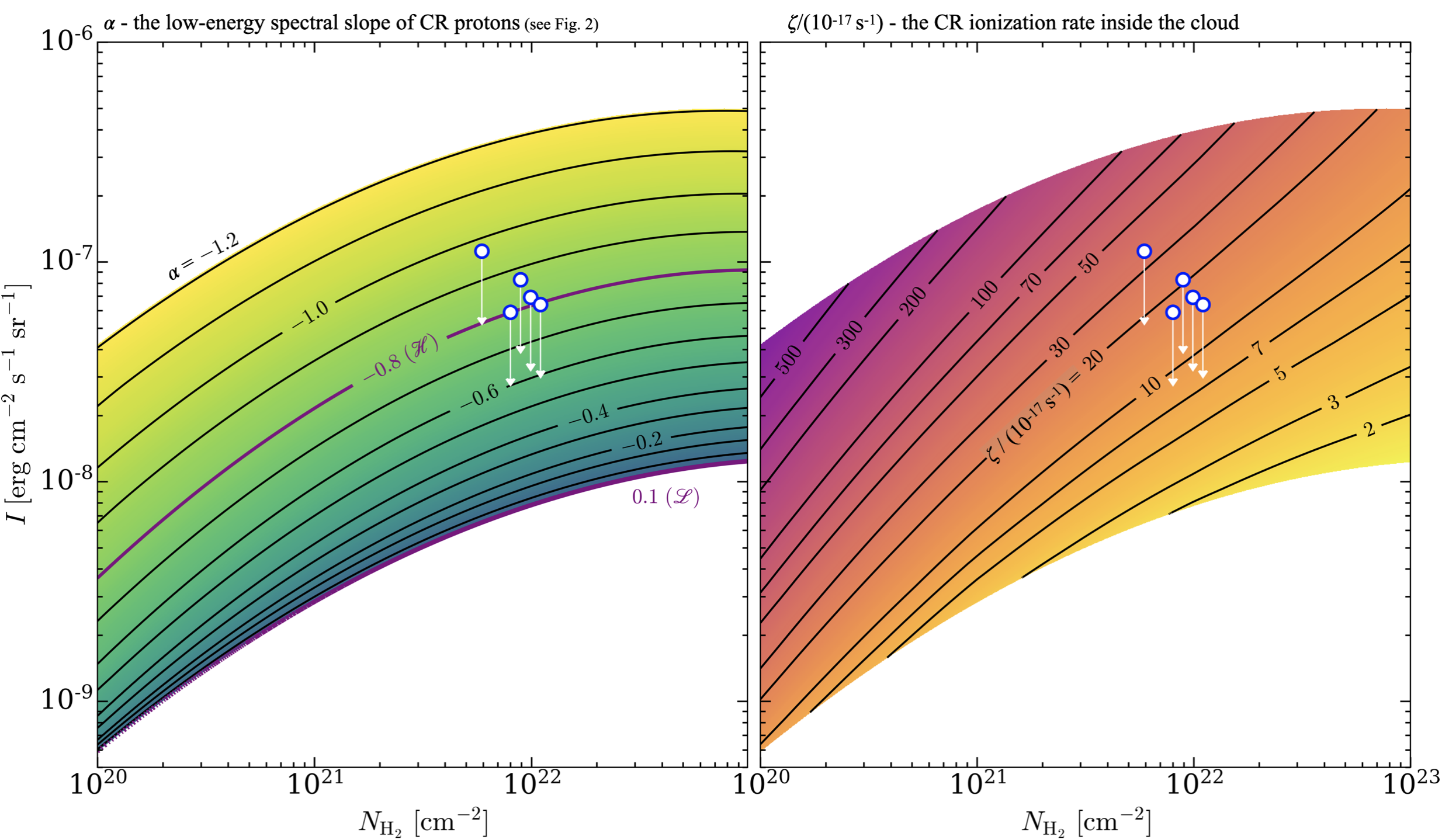}
    \caption{The $(1-0)$S(0) line brightness as a function of the cloud H$_2$ column density for different models of the interstellar  CR proton spectrum impinging upon the cloud (see Fig.~\ref{fig: ISM spectra}).
    Contours: the CR spectral slope $\alpha$ (left), and the (attenuated) CR ionization rate inside the cloud (right). 
    The five markers are our 3$\sigma$ upper limits on the $(1-0)$S(0) line, which translate into upper limits on $\zeta$ and lower limits on $\alpha$.
    }
    \label{fig: zeta_limits}
\end{figure*}

 \subsection{The CR spectrum - numerical model}
 \label{sub: CRIR limit numerical}
We use a detailed numerical model to connect the  CR ionization and excitation rate in the cloud interior
to the initial spectrum of low-energy CR protons that is impinging the cloud on its boundary.
This allows us to convert the upper limit on the $(1-0)$S(0) line brightness into a constraint on the low-energy interstellar CR proton spectrum.
 We account for energy losses due to H$_2$ ionization, dissociation, excitation, and momentum transfer, and calculate the modulation of the CR spectrum (of both primary and secondary CRs) as CRs propagate into a cloud. We derive the 
 resulting H$_2$ excitation rate and the brightness of the H$_2$ rovibrational emitted lines for different interstellar CR spectra.
 For more details, see Padovani et al.~(2022, in prep.), hereafter P22.

For the  interstellar CR proton spectrum, we consider a continuous set of models that are characterized by  their low-energy spectral slope, $\alpha$. 
An example of three models is shown in Fig.~\ref{fig: ISM spectra} (left panel).
At energies $E \gtrsim 1$ GeV the proton spectrum is constrained by AMS-02 \citep{Aguilar2015}
For $E = 3-300$ MeV, the proton spectrum has been observed by the two Voyager spacecrafts 
\citep{Cummings2016, stone2019}. However, it is not clear whether Voyager is probing a representative interstellar CR field. First, the magnetic field direction measured by the Voyager probes did not show the change expected if they were beyond the influence of solar modulation \citep{Gloeckler2015, Padovani2018a}. 
Second, even if the probes are not under the influence of solar modulation, they are still in the local bubble.
Consequently, there is a substantial uncertainty about the low-energy proton spectrum at $E \lesssim 1$ GeV,
and hence we consider an array of models with different $\alpha$ values.
For the CR electron flux we follow P22 (their Eq.~2 and Table 2).

The total energy density, $n_{\rm CR}$, for the various models is indicated in Fig.~\ref{fig: ISM spectra} (left). 
The steeper spectra (smaller $\alpha$ values) have higher CR fluxes, and consequently, $\zeta$ increases as $\alpha$ decreases  - see Fig.~\ref{fig: ISM spectra}, upper-right panel. 
Fig.~\ref{fig: ISM spectra} also shows that for a given spectrum (fixed $\alpha$), $\zeta$ decreases with increasing cloud column $N_{\rm H_2}$. This is because the CRs lose energy as they propagate into the cloud and the CR flux at low energies 
decreases with $N_{\rm H_2}$ (see P22, Fig.~4).
The lower-right panel of Fig.~\ref{fig: ISM spectra} shows the excitation to ionization rate ratio for the two states $v=1$, $J=0$ and $J=2$, which dominate the CR excitation of H$_2$ (\citetalias{Bialy2020}).
This ratio varies with $N_{\rm H_2}$ and $\alpha$, however the variations are rather mild. 
Summing up the excitation of both levels, we obtain $\phi \equiv \zeta_{\rm ex}/\zeta = 3.6$ to 4.8.
Excitation to $v \geq 2$ levels increase $\zeta_{\rm ex}$ by $\approx 10$ \% giving $\phi = 4 - 5.3$, in reasonable agreement with the value used by \citetalias{Bialy2020}.


Using our CR propagation model we have generated a lookup plot, Fig.~\ref{fig: zeta_limits}, that predicts the interstellar proton CR spectral slope $\alpha$ (i.e., the models shown in Fig.~\ref{fig: ISM spectra}), and the CR ionization rate in the cloud interior, $\zeta(N_{\rm H_2})$, given a measurement of the $(1-0)$S(0) line brightness and the cloud's column $N_{\rm H_2}$ (see P22 for additional lines).
For a given $\alpha$ value, the integrated line intensity increases with $N_{\rm H_2}$ in the optically thin limit, and then saturates at $N_{\rm H_2} \approx 10^{22}$ cm$^{-2}$ as the cloud becomes optically thick due to dust absorption (see Appendix \ref{app: analytic model CR attenuation}).
At a given $N_{\rm H_2}$, $\zeta$ increases with decreasing $\alpha$ because the lower $\alpha$ models correspond to higher CR fluxes.

The five markers show our 3$\sigma$ upper limits on the $(1-0)$S(0) line for G150, G150p2, G157, G157p2 and G163.
Correspondingly, the lower limit on $\alpha$ is within -0.87 and -0.67 for these clouds, and the upper limit on $\zeta$ is within $1.0$ and $2.6 \times 10^{-16}$ (see Table \ref{table}).
The $\zeta$ values are in excellent agreement with the analytic model (\S \ref{sub: CRIR limit analytic}), and are in agreement with the general range of $\zeta$ values in the literature, deduced via absorption spectroscopy of various molecules (see P22 for a comprehensive comparison).

\section{Future prospects for JWST}
\label{sec: JWST}

Despite our long integration time the H$_2$ rovibrational lines were not detected in any of our 6 targets.
This is because at this wavelength range ($\approx 2-3$ $\mu$m) the spectrum is contaminated by thermal emission, absorption, and strong skylines from the atmosphere. 
We estimate JWST/NIRSpec's expected sensitivity for line detection.
First, let us focus on the $(1-0)$O(2) line at $\lambda=2.63$ $\mu$m. This line is blocked for ground-based observations, but is predicted to be the strongest line for CR excitation (\citetalias{Bialy2020}).  
The O(2) line emission from a cloud with column $N_{22}=1$ illuminated by a CR spectrum with slope $\alpha=0.1$ (for which $\zeta \approx 10^{-17}$ s$^{-1}$), is $I \approx 3 \times 10^{-8}$ erg cm$^{-2}$ s$^{-1}$ sr$^{-1}$ (P22).
Using JWST's exposure time calculator (ETC) with $t_{\rm exp}=1.25$ hrs
we obtain a SNR $= 1.24$ per shutter.
Integrating over the $365 \times 2=730$ shutters along the spatial direction 
gives 
${\rm SNR} \approx 33.5$.
 Importantly, the line brightness that we assumed corresponds to a proton spectrum with $\alpha=0.1$ - this is the spectrum with the shallowest slope, the lowest $\zeta$ value, and the faintest emission among all models (see Fig.~\ref{fig: zeta_limits}, and Fig.~8 in P22 for additional H$_2$ lines).
 Thus, in practice, JWST will be sensitive to the entire range of possible CR models, and will be able to robustly constrain the proton spectrum at low energies.

Another advantage of JWST's high sensitivity is that it will allow the discrimination of various H$_2$ excitation mechanisms, including excitation by CRs, UV pumping and chemical excitation from H$_2$ formation.
As discussed in \citetalias{Bialy2020}, these excitation processes exhibit different line ratios. 
For example, the line ratio $\eta \equiv I_{\rm (1-0)S(1)}/I_{\rm (1-0)S(0)}$: for pure UV excitation $\eta \approx 2$ \citep{Black1987, Sternberg1988}, for pure H$_2$ formation $\eta \approx 3.5 - 5.6$ \citep{LeBourlot1995}, whereas for pure CRs $\eta \approx 0.04$ or lower (\citetalias[][]{Bialy2020}). 
Thus, measured line ratios may be used to determine the relative 
importance of each excitation process and the individual values of 
the CR and the FUV radiation fluxes  (see Appendix \ref{app: UV vs CRs} for strategies for separating FUV excitation vs CR excitation).
To evaluate the ability of JWST to detect the fainter lines, we consider the $(1-0)$S(1) line ($\lambda =2.12$ $\mu$m), excited by the mean FUV interstellar radiation field \citep{Draine1978, Bialy2020c}, for which $I = 1.8 \times 10^{-8}$ erg cm$^{-2}$ s$^{-1}$ sr$^{-1}$ (Eqs.~8-9, \citetalias{Bialy2020}; this analytic value is in good agreement with the numerical results discussed in Appendix \ref{app: UV vs CRs}).
Using JWST/NIRSpec's ETC with $t_{\rm exp}=1.25$ hrs we obtain a SNR $=1.22$ per shutter and an integrated SNR $\approx 33$ (over all shutters).
Similarly, other H$_2$ rovibrational lines may be also robustly detected.

Given a detection, we may also restrict the shutter integration over a smaller number of shutters.
This will sacrifice the SNR but will allow to derive the CR ionization rate in different positions along the observed cloud, allowing to obtain, for the first time, the $\zeta$ gradient across a starless core with a very high spatial resolution of\footnote{We used NIRSpec's shutter angular size $0.53''$ at a distance $d=500$ pc} $\sim 10^{-3}$ pc.

X-rays have a similar effect on H$_2$ 
excitation as CRs. This is because in both case, the H$_2$ is excited mainly by the secondary electrons (see P22 for a comparison of primary vs secondary excitation). Thus, when constraining CR properties, it is important to ensure that the observed cloud is starless. 
On the other hand,  H$_2$ emission lines may be used to constrain the properties of the X-ray irradiation, if the targeted clouds are specifically chosen to be in the vicinity of known X-ray sources, e.g., near X-ray binaries or active galactic nuclei.



\section{Conclusions}
\label{sec: conclusions}
Utilizing the ``direct method" H$_2$ analysis (including the \citetalias{Bialy2020}'s analytic model and P22's numerical model) and deep NIR spectroscopy of several dense clouds, we placed upper limits on the ionization rate and on the spectral slope of low-energy CR protons, ruling out steep CR spectral models.
While ground based observations cannot detect the H$_2$ rovibrational lines, JWST/NIRSpec will be able to efficiently detect them, and thus to determine the relative roles of CR versus UV excitation, and to constrain the otherwise elusive low-energy CR spectrum.  
Extending this observation to a sample of clouds located in different proximity to potential CR sources, as well as in different positions within a molecular cloud, will allow to constrain fluctuations in the CR ionization rate, the source of low-energy CRs and the CR propagation process.

\begin{acknowledgements}
Observations reported here were obtained at the MMT Observatory, a joint facility of the Smithsonian Institution and the University of Arizona, through the program ``SAO-12-20c Using Molecular Clouds as Cosmic Ray Detectors", PI: S. Bialy. This paper uses data products produced by the OIR Telescope Data Center, supported by the Smithsonian Astrophysical Observatory. S.~Bialy acknowledges support from the Institute for Theory and Computations at the Harvard-Smithsonian Center for Astrophysics, and from the Center for Theory and Computations at University of Maryland, College Park.
S.~Belli acknowledges support from the Clay Fellowship. 
We thank Igor Chilingaryan, Sean Moran, and Bryan McLeod for helpful discussions on of the MMT data and pipeline.
\end{acknowledgements}


\begin{appendix}

\section{Data reduction and noise derivation}
\label{app: data reduction}
\begin{figure*}
    \centering
    \includegraphics[width=1\textwidth]{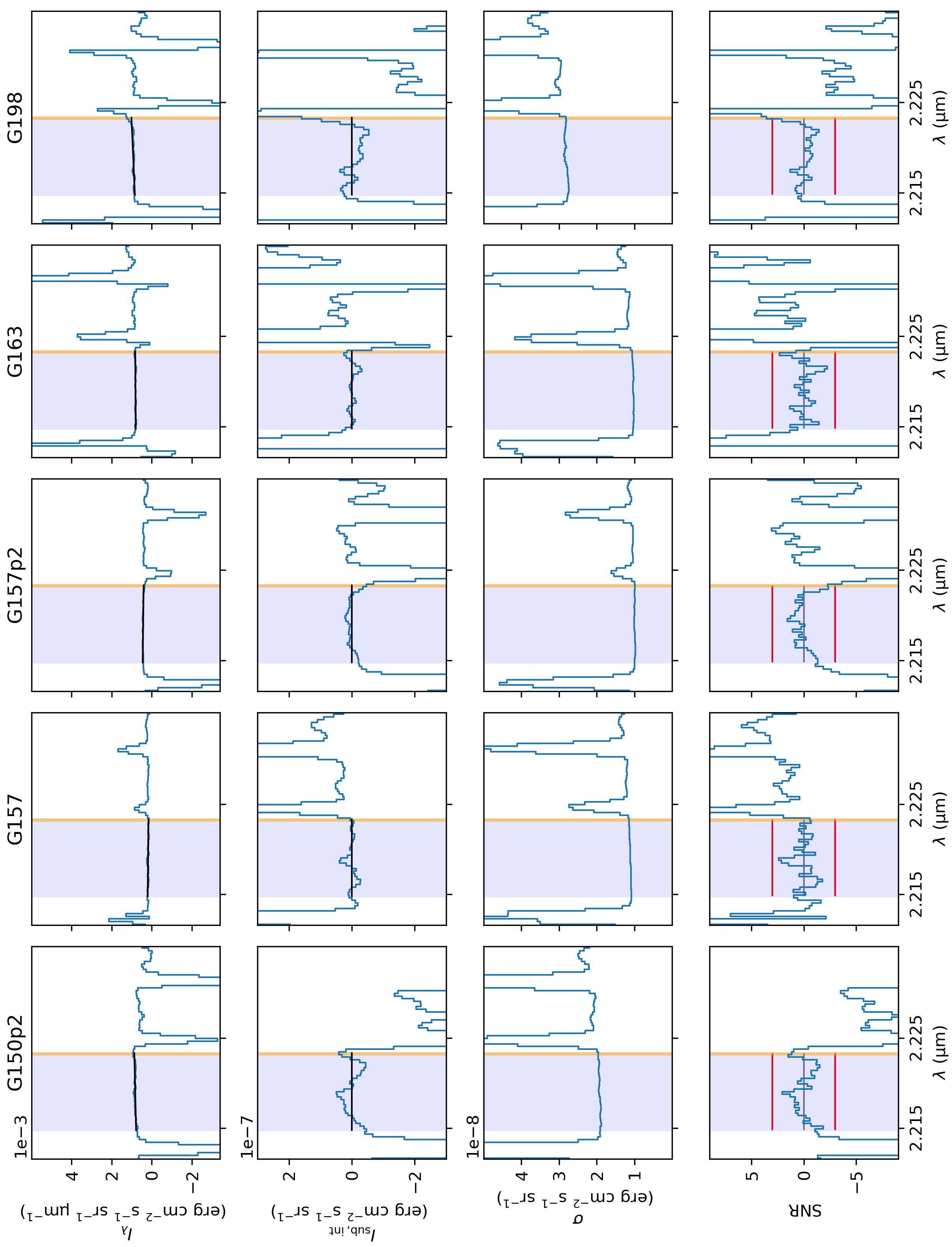}
    \caption{Same as Fig.~\ref{fig: spectrum}, but for the rest of our targets.
    }
    \label{fig: spectrum all}
\end{figure*}

The observations were obtained with the MMIRS instrument on the MMT over several nights between November 2020 and January 2021, using the K3000 grism with a 0.4 arcsec slit width, yielding a spectral resolution $R\sim3000$.
(using a wider slit would increase the amount of light reaching the instrument, but would also broaden the sky emission lines). 
The MMIRS slit is 7 arcmin long, which is a good match to the angular size of the targets. 
The slit was generally placed on the peak of the column density distribution. 
To improve sky subtraction we alternated object frames to sky frames, which were obtained by moving the telescope by 30 arcmin along the direction of the slit. Each target was observed in 5-minute frames for a total of approximately 180 min on source and 180 min on sky. 

\subsection{Data reduction}
The observations were reduced with the MMIRS data reduction pipeline \citep{Chilingarian2015}. Since the emission from the cloud covers the entire extent of the longslit, it is not possible to model and subtract the sky emission from the data, and we skipped the sky subtraction step that uses the \citet{Kelson2003} method. Instead, we rely only on the difference of source-sky frames to suppress the sky emission.

After masking bad pixels and stellar traces, which usually account for 20-30\% of the pixels in the 2D reduced spectra, we stack the spectrum of each target along the slit and obtain the 1D spectrum. We apply a theoretical flux calibration based on the telescope collecting area, angular size of the observed region, exposure time, expected efficiency of the instrument, and we convert the spectrum from counts to flux density per unit solid angle (erg cm$^{-2}$ s$^{-1}$ $\mu$m$^{-1}$ sr$^{-1}$).
We compare this flux calibration to that obtained using observations of standard stars, and we conclude that the overall flux calibration is reliable within a factor of two.

\subsection{Deriving the noise and placing upper limits on line brightness}
Our observations span the wavelength range $1.890-2.509~\mu$m, which contains several H$_2$ transitions. However, most of these are either expected to be too weak to be detected, or strongly affected by sky emission and absorption. This leaves two lines of interest: $(1-0)$S(0) and $(1-0)$S(1); we focus on the former since for CR excitation in cold molecular clouds, (1-0)S(0) is expected to strongly dominate over (1-0)S(1) (see \citetalias{Bialy2020}, Table 1). 
We still checked for $(1-0)$S(1) but did not detect any emission.

The upper panels of Fig.~\ref{fig: spectrum} and Fig.~\ref{fig: spectrum all} show a section of the 1D spectrum in the vicinity of the $(1-0)$S(0) line ($\lambda=2.22 \mu$m, denoted by the yellow strip) for all of our targets. 
The strong features, including the high peak a few pixels redward of the (1-0)S(0) line, are skylines
which we use to explicitly verify that the wavelength calibration is very precise.
We identify a region around the (1-0)S(0) line that is devoid of skylines, marked as the blue strip, hereafter the ``good $\lambda$ range".
For this wavelength range we fit a linear function, shown as the black line, and subtract it from the spectra.
Since the H$_2$ emission lines are very narrow (the instrument resolution is $\approx 100$ km/s. In contrast,  thermal broadening and turbulent broadening for typical cold clouds are $\approx 0.2$ km/s and $\approx 2$ km/s), the line, if observed, would be broadened to the instrument's spectral resolution, $\Delta \lambda = \lambda/R \approx 7.4 \times 10^{-4}$ $\mu{\rm m} \approx 2.5$ pixels.
We have also verified this value by measuring the width of nearby skylines in the observed spectrum.
Thus, to search for a line emission, we integrate the subtracted spectrum over $\Delta \lambda$.
We denote this subtracted and integrated spectrum $I_{\rm sub, int}$, and show it in the 2$^{\rm nd}$ rows of 
Figs.~\ref{fig: spectrum}, \ref{fig: spectrum all}.
 The $(1-0)$S(0) emission line is not detected for any of the targets.

To place an upper limit on the (1-0)S(0) line brightness we evaluate the noise:
\begin{enumerate}
    \item 
    In the 2D spectra, we sum in quadrature the error on each pixel, along the spatial dimension (excluding the masked pixels).
    The error per pixel is provided by MMIRS's pipeline, based on the standard deviation (STD) in the pixel flux across individual time frames. We also compared it with the signal's STD across the spatial and wavelength directions, and also with a theoretical error estimate based on calculations of the Poisson noise and readout noise and found a good agreement.
    We integrate these errors over $\Delta \lambda$ and denote the result $\sigma_{\rm pipe}$.
    \item 
    Comparing $\sigma_{\rm pipe}$ with the fluctuations of $I_{\rm sub, int}$ along the wavelength direction, we find that the $\sigma_{\rm pipe}$ values are too optimistic.
    For example, the calculated STD of $I_{\rm sub, int}$ within the ``good $\lambda$ range" (the blue strip in Fig.~\ref{fig: spectrum}) for G150 is $1.6 \times 10^{-8}$ erg cm$^{-2}$ s$^{-1}$ sr$^{-1}$, a factor of 2.1 higher than $\langle \sigma_{\rm pipe} \rangle$ (average over the good $\lambda$ range).
    Thus, for G150, we correct the $\sigma_{\rm pipe}$ values by multiplying them by 2.1.   
    For the rest of our targets, the correction factors range within 1.5 to 2.7.
    This correction ensures that the STD of $I_{\rm sub, int}$ equals the mean $\sigma$ values, as it should.
    We denote the corrected errors $\sigma$. 
\end{enumerate}
  
The corrected errors, $\sigma$, are shown in the 3$^{\rm rd}$ rows of Figs.~\ref{fig: spectrum}, \ref{fig: spectrum all}.
The 4$^{\rm th}$ rows show the SNR.
At the (1-0)S(0) line wavelength (orange strip), ${\rm SNR }<3$, and thus we claim a non-detection.
More generally, within the ``good $\lambda$ range", there are fluctuations but overall $\| {\rm SNR}\| < 3$, 
giving confidence in our evaluation of the spectrum and error corrections.

We use the 3$\sigma$ noise level at the (1-0)S(0) line wavelength to place an upper limit on the (1-0)S(0) line brightness.
We further multiply the 3$\sigma$ values by a factor of 2 to account for the flux calibration uncertainty (see above).
This gives us robust upper limits on the (1-0)S(0) line brightness, which we report in Table \ref{table}, and show in Fig.~\ref{fig: zeta_limits}.

\section{Analytic model for the case of cosmic ray attenuation}
\label{app: analytic model CR attenuation}
In this appendix we discuss a generalization for Eq.~(\ref{eq: I-zeta relation}) for the case of a non-constant $\zeta$, i.e., a case where CR energy losses are taken into account and thus result in a $\zeta$ that decreases with cloud depth.
This case was also discussed in \citetalias{Bialy2020} (see their Methods section).
However, as we discuss below, \citetalias{Bialy2020}'s conclusion that for a varying CR rate, Eq.~(\ref{eq: I-zeta relation}) still holds if $\zeta N_{\rm H_2}$ is replaced with $\int \zeta \mathrm{d}N_{\rm H_2}$ is only correct in the case of optically thin gas ($\tau \ll 1$). As we show for a varying $\zeta$ and an optically thick gas, an analytic solution can still be obtained, but it is more involved.

The contribution to the line emission from an infinitesimal slab, ${\rm{d}}N_{\rm H_2}$, that is excited by CR particles and their produced secondary electrons, may be written as:
\begin{equation}
\label{eq: dI}
 {\rm{d}}I_{ul} = \frac{1}{4 \pi} {\rm{d}}N_{\rm H_2}  \zeta_{\rm ex}(N_{\rm H_2})  p_{u} \alpha_{(u)l} \mathrm{e}^{-\tau} 
\end{equation}
where $N_{\rm H_2}$ is the integrated column from cloud edge to the position of the infinitesimal slab, $\tau = \sigma_{d, {\rm H_2}} N_{\rm H_2}$ is the optical depth and  and $\sigma_{d, {\rm H_2}} \approx 0.9\times 10^{-22}$ cm$^{2}$ is the dust absorption cross section per H$_2$ molecule, at the wavelength of interest
(\citealt{Draine2003}, \citetalias{Bialy2020}). Dust absorption dominates the opacity for the considered lines. The meaning of the rest of the symbols is as described in \S \ref{sub: CRIR limit analytic}.

For a cloud of total H$_2$ column density $N_{\rm H_2}$, we integrate Eq.~(\ref{eq: dI}) and obtain the emitted line brightness:
\begin{equation}
\label{eq: I integral}
     I_{ul} = \int {\rm{d}}I_{ul} = \frac{1}{4 \pi} p_{u} \alpha_{(u)l} \int_0^{N_{\rm H_2}} \zeta_{\rm ex}(N_{\rm H_2}^{\prime})  \mathrm{e}^{-\tau} {\rm{d}} N_{\rm H_2}^{\prime} \ .
\end{equation}
This integral may be solved under some simplifying assumptions.

\subsection{Constant CR excitation rate: both optically thin and thick regimes}
For a constant CR excitation rate (i.e., where $\zeta_{\rm ex}$ is not a function of $N_{\rm H_2}$) we can pull $\zeta_{\rm ex}$ out of the integral and we get
\begin{align}
\label{eq: I integral const zeta}
     I_{ul} &= \frac{1}{4 \pi}p_{u} \alpha_{(u)l} \zeta_{\rm ex}  \int_0^{N_{\rm H_2}} \mathrm{e}^{-\tau} {\rm{d}} N_{\rm H_2}^{\prime} \nonumber \\
     &= \frac{1}{4 \pi}p_{u} \alpha_{(u)l} \zeta_{\rm ex} g N_{\rm H_2}
\end{align}
where 
\begin{equation}
    g \equiv \frac{1-\mathrm{e}^{-\tau}}{\tau}  = \frac{1-\mathrm{e}^{-0.9 N_{22}}}{0.9 N_{22}} \ ,
\end{equation}
and $N_{22}\equiv N_{\rm H_2}/(10^{22} \ {\rm cm^{-2}})$.
This converges with Eq.~(\ref{eq: I-zeta relation}). The $g$ factor includes the optical thickness effect.
For small column densities ($\tau \ll 1$), $g \rightarrow 1$ and  $I_{ul} \propto N_{\rm H_2}$ as expected for optically thin emission.
As $N_{\rm H_2}$ increases, $I_{ul}$ increases until at sufficiently large 
columns ($\tau \gtrsim 1$), $g \rightarrow \tau^{-1} = (0.9 N_{22})^{-1}$. 
In this limit $I_{ul}$ saturates and becomes independent of $N_{\rm H_2}$. This is the optically thick regime.

\subsection{Non constant CR excitation rate: the optically thin regime}
In practice, the CR excitation and ionization rates are expected to vary with cloud depth, due to CR energy losses \citep{Padovani2009}.
For a non constant CR excitation rate, but assuming the optically thin regime $\tau<1$, we have
\begin{equation}
\label{eq: I integral optically thin limit 0}
I_{ul} = \frac{1}{4 \pi} p_{u} \alpha_{(u)l} \int_0^{N_{\rm H_2}}   \zeta_{\rm ex}(N_{\rm H_2}^{\prime}) {\rm{d}}N_{\rm H_2}^{\prime}  \ .
\end{equation}
While generally the functional form of $\zeta_{\rm ex}(N_{\rm H_2})$ may be complex, if we are interested in sufficiently small column densities, $N_{\rm H_2} \lesssim 10^{24}$ cm$^{-2}$, $\zeta_{\rm ex}(N_{\rm H_2})$ may be approximated as a power-law $\zeta_{\rm ex} = \zeta_0 (N_{\rm H_2}/N_0)^{-a}$ where $a$ is typically within the range $(0,1)$, and its exact value depends on the interstellar CR proton spectrum  (see Fig.~5 in P22). We get
\begin{align}
\label{eq: I optically thin limit}
 I_{ul} = 
    \frac{1}{4 \pi} p_{u} \alpha_{(u)l}  \frac{\zeta_{0} N_0}{1-a} \left( \frac{N_{\rm H_2}}{N_0}\right)^{1-a} 
    = 
 \frac{1}{4 \pi} p_{u} \alpha_{(u)l}  \zeta_{\rm ex}(N_{\rm H_2}) N_{\rm H_2} \frac{1}{1-a}
 \ .
\end{align}
where $\zeta_{\rm ex}(N_{\rm H_2})$ is the CR excitation rate inside the cloud interior.
We see that in the optically thin limit, and for a non-constant CR excitation rate, we still obtain an equation similar to Eq.~(\ref{eq: I-zeta relation}) (with $g=1$ by definition as we assumed the optically thin regime), but with a correction factor $1/(1-a)$, which is typically of order unity.
The powerlaw $a$ depends on the interstellar CR proton spectrum.
For example, for the column density range $N_{\rm H_2}=10^{20}-10^{23}$ cm$^{-2}$, $a\approx 0.38$ for the $\mathcal{H}$ proton spectrum, and $a\approx 0.06$ for the $\mathcal{L}$ proton spectrum (see P22, Fig.~4).
For these $a$ values, the correction factor is 1.6 and 1.1, respectively.

\subsection{Non constant CR excitation rate: the general case}
For the general case where the CR rate varies with cloud depth, and the cloud is not optically thin,
the integral may still be solved if we assume a power-law form for $\zeta_{\rm ex}(N_{\rm H_2})$. 
We get
\begin{align}
\label{eq:  I full integral zeta powerlaw}
    I_{ul} &= \frac{1}{4 \pi}p_{u} \alpha_{(u)l} \int_0^{N_{\rm H_2}} \zeta_{\rm ex}(N_{\rm H_2}^{\prime})  \mathrm{e}^{-\tau} {\rm{d}} N_{\rm H_2}^{\prime}
    \nonumber \\
    &= \frac{1}{4 \pi}p_{u} \alpha_{(u)l} \zeta_0  \frac{\tau_0^a}{\sigma_{d, {\rm H_2}}} \int_0^{\tau} x^{-a} \mathrm{e}^{-x}  {\rm{d}} x \nonumber \\
    &=  \frac{1}{4 \pi}p_{u} \alpha_{(u)l} \zeta_0  \frac{\tau_0^a}{\sigma_{d, {\rm H_2}}} \gamma(1-a, \tau)
\end{align}
where we defined $\tau_0 \equiv \sigma_{d, {\rm H_2}} N_0$, and where $\gamma$ is the lower incomplete gamma function.
In the optically thin limit ($\tau \ll 1$), the gamma function $\gamma(1-a, \tau) \rightarrow \tau^{1-a}/(1-a)$ and Eq.~(\ref{eq:  I full integral zeta powerlaw}) then approaches the solution in Eq.~(\ref{eq: I optically thin limit}) as it should.

\section{H$_2$ photo-excitation versus cosmic-ray excitation}
\label{app: UV vs CRs}

\subsection{FUV and CR excitation - physical processes}
Although most of the cloud volume is devoid of FUV radiation due to efficient dust absorption, the  ambient interstellar FUV radiation field will lead to H$_2$ excitation in a thin layer at the cloud envelope, i.e., the photo-dominated region (PDR). 
Here, FUV photons within the Lyman-Werner band ($11.2-13.6$ eV) populate the excited electronic states of H$_2$, which then radiatively decay, populating the excited rovibrational states of the ground electronic state. These rovibrational states (denoted $vJ$), radiatively decay producing a rich spectrum of emission lines in the NIR.
This FUV excitation produces lines with intensities that may be comparable to those produced by CR excitation. 
Whether the line emission is dominated by FUV excitation or CR excitation depends on the relative intensity of the FUV radiation field to the CR ionization rate, $I_{\rm UV}/\zeta$, and on the specific line considered (see Eqs.(10-11) in \citetalias{Bialy2020}), where $I_{\rm UV}=F_{\rm UV}/F_{\rm UV,0}$ is the FUV interstellar radiation flux normalized to the solar neighborhood value $F_{\rm UV,0}=2.7 \times 10^{-3}$ erg cm$^{-2}$ s$^{-1}$ \citep{Draine1978, Bialy2020c}

\begin{figure*}
    \centering
    \includegraphics[width=0.9\textwidth]{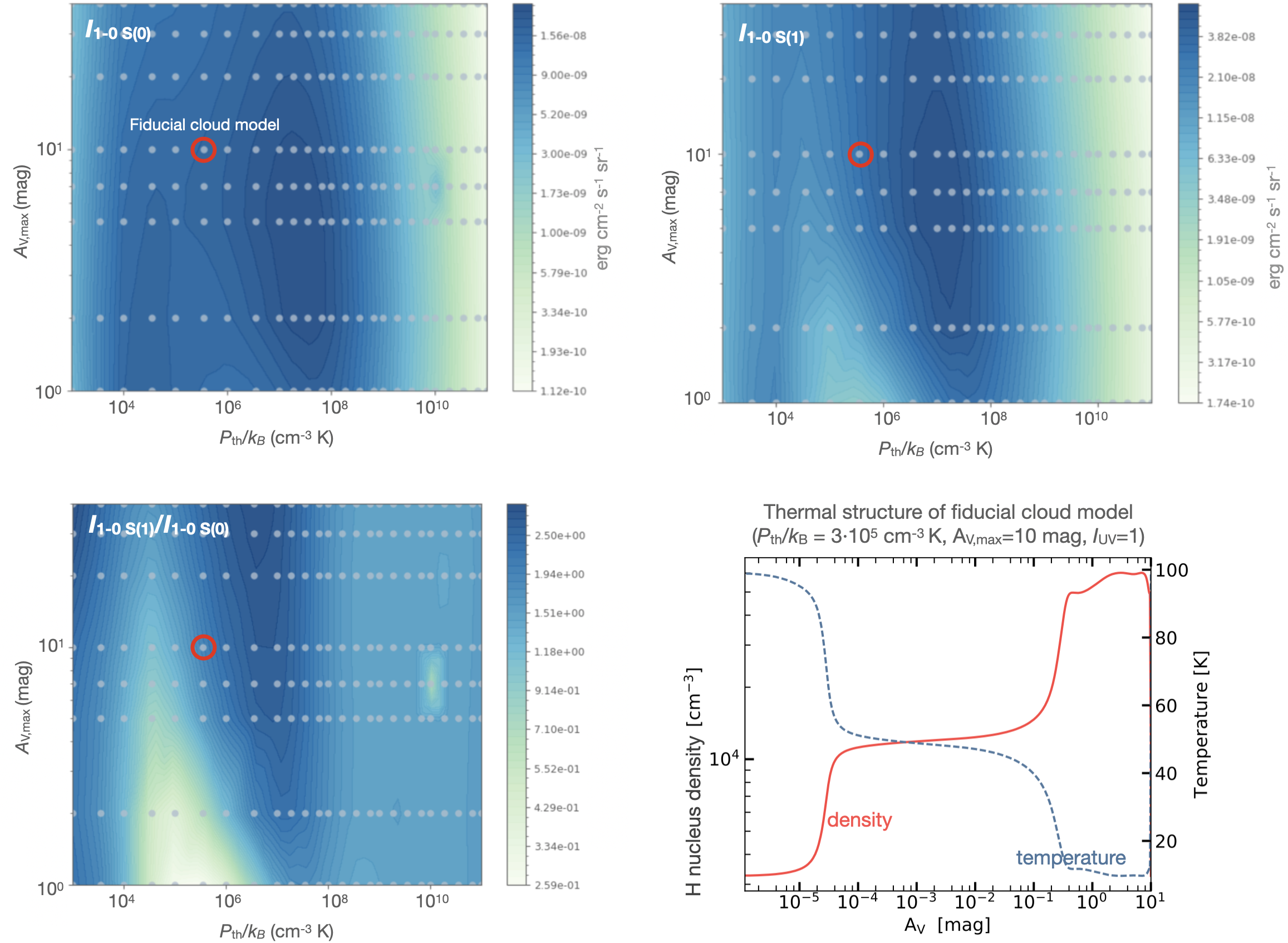}
    \caption{The (1-0)S(1) and (1-0)S(0) emission brightness (top), and their ratio (bottom-left) for pure FUV photo-excitation as computed by the isobaric MEUDON PDR models \citep{LePetit2006}.
    The line brightness are presented in the parameter space 
    of the cloud's total visual extinction, $A_{\rm V, max}$, versus the cloud's thermal pressure, $P_{\rm th}$.
    The points are the locations where the PDR models were computed. The red circle is the fiducial model which represents a typical cloud for our observations, for which we also show the thermal structure of the cloud (bottom-right).}
    \label{fig: pdr models}
\end{figure*}

Interestingly, FUV excitation and CR excitation produce different excitation pattern of the H$_2(vJ)$ levels, and thus predict different ratios for the various H$_2$ emission lines.
This is due to two reasons.
First is because the excitation processes are physically different.
For FUV excitation, the H$_2(vJ)$ are populated through radiative cascade from the excited electronic states (see above), whereas in the case of CRs, direct impact excitation is important. In this latter process, the secondary electrons produced by CR ionization interact directly with the H$_2$ nuclei efficiently exciting its first vibrational states (\citealt{Gredel1995}, P22). This results in very high ratios for lines emitted from H$_2(v=1)$. For example, the emission of (1-0)S(0) is stronogy enhanced, compared to lines from higher vibrational levels, e.g., (2-1)S(0). This is different from FUV excitation which efficiently excites a large array of levels, including those with high $v$ numbers.

The second reason for the different line ratios for FUV versus CRs, is that the H$_2$ excitation 
takes place in different regions of the cloud for the two processes, with significant differences in the gas temperature.
The FUV excitation occurs at the cloud outer PDR layer. Here the gas is efficiently heated by the FUV radiation (including: photoelectric heating, H$_2$ FUV-pumping heating, and H$_2$ formation heating). The balance between heating and cooling results in gas temperatures of order $100$ K. With increasing cloud depth, the FUV radiation is absorbed by dust, and the heating rate decreases. In the deep cloud interiors the H$_2$ gas is colder and denser, with typical temperatures of order 10 K. 
These temperature differences result in different ortho-to-para H$_2$ ratios, such that the ortho-to-para ratio in the cloud interior is significantly lower compared to the PDR.
Thus, the FUV excitation results in the emission of both odd and even lines, with comparable intensities, e.g., the (1-0)S(1) and (1-0)S(0), whereas for CR excitation only the H$_2$ even $J$ states are predominantly excited.
For example, for CR excitation, the ratio $\eta \equiv I_{\rm (1-0)S(1)}/I_{\rm (1-0)S(0)}$ is predicted to be very low: for $T=30$ K, $\eta = 0.04$, and it further decreases with decreasing temperature (\citetalias{Bialy2020}), whereas for FUV excitation in the warmer PDR, $\eta$ is typically of order unity \citep{Black1987, Sternberg1988, Sternberg1989b}.

One may claim that the line ratio is thus not tracing the excitation mechanism (CR vs FUV) but simply the gas temperature. 
However, in practice, the gas thermal structure is not arbitrary, but is {\it controlled} by the intensities of FUV and CRs in the cloud. The gas in the cloud envelope (PDR), is warmer because it is efficiently heated by the FUV radiation.
These same FUV photons are also those that excite the H$_2$ in the PDR. In the deep cloud interior, the gas is colder because the FUV radiation is excluded. Here CRs both excite the H$_2$ and control the gas temperature 
(i.e., through ionization and chemical heating; \citealt{Glassgold2012}).

\subsection{Strategies for constraining the contributions of FUV and CR excitation}
In a realistic observation, the various H$_2$ lines and their ratios are influenced by both the FUV excitation (in the PDR) and the CR excitation (in the cloud interior).
Thus, any considered line ratio would have an intermediate value between the ``pure-FUV" expected value and the ``pure-CR" value. The value depends on the FUV and CR intensities, $I_{\rm UV}$, $\zeta$.

As an example, let us estimate the (1-0)S(1) and (1-0)S(0) line emissions produced by the combined effect of FUV and CR excitation, and the resulting ratio of the two lines for typical starless cores like those observed in the present paper. 
For the contribution of FUV excitation we utilize the MEUDON PDR model results \citep{LePetit2006, LeBourlot1995, Bron2014}\footnote{\href{https://ism.obspm.fr/ismdb.html}{https://ism.obspm.fr/ismdb.html}}.
For our fiducial model, we assume $I_{\rm UV}=1$, $\zeta=10^{-16}$ s$^{-1}$, a total cloud visual extinction $A_{\rm V, max}=10$ mag ($N_{\rm H_2}\approx 10^{22}$ cm$^{-2}$), and a cloud thermal pressure $P_{\rm th}/k_B=3 \times 10^5$  cm$^{-3}$ K (corresponding to inner density and temperature $n_{\rm H_2} \approx 3 \times 10^4$ cm$^{-3}$, $T \approx 10$ K). 
We focus on isobaric models as they obey force equilibrium across the cloud layers (i.e., the pressure is constant as a function of cloud depth).
In Fig.~\ref{fig: pdr models} we present contour plots showing the line emission of (1-0)S(1), (1-0)S(0), and their ratio, as obtained by the MEUDON PDR model (in which the H$_2$ is excited only by FUV) in the $A_{\rm V, max}-P_{\rm th}$ parameter space. 
The fiducial model is highlighted by the red circles. 
The thermal and density structure for the fiducial model is presented in the lower-right panel.

For the fiducial model, pure FUV excitation results in $I_{\rm (1-0)S(0)}^{\rm FUV} = 1.6 \times 10^{-8}$ erg cm$^{-2}$ s$^{-1}$ sr$^{-1}$ and  $I_{\rm (1-0)S(1)}^{\rm FUV} = 2.4 \times 10^{-8}$ erg cm$^{-2}$ s$^{-1}$ sr$^{-1}$, and their ratio is $\eta^{\rm FUV}=1.5$.
For most of the parameter space, these values are only weakly dependent on the exact values of  $A_{\rm V, max}$ and $P_{\rm th}$.
For the CR contribution we use Eqs.~(3-5) in \citetalias{Bialy2020} with
$f=0.16$ appropriate for (1-0)S(0) (Table 1 in \citetalias{Bialy2020}), $N_{\rm H_2}=9.4 \times 10^{21}$ cm$^{-2}$ (equivalent to $A_{\rm V, max}=10$ mag), and $\zeta=10^{-16}$ s$^{-1}$. We get  $I_{\rm (1-0)S(0)}^{\rm CR}=3.6 \times 10^{-8}$ erg cm$^{-2}$ s$^{-1}$ sr$^{-1}$. For (1-0)S(1), CR excitation is negligible compared to FUV excitation.
Accounting for both contributions, we get the (1-0)S(0) and (1-0)S(1) line intensities $(I_{\rm (1-0)S(0)}, I_{\rm (1-0)S(1)}) = (5.2, 2.4) \times 10^{-8}$ erg cm$^{-2}$ s$^{-1}$ sr$^{-1}$, 
and the ratio is $\eta = 0.46$.
Thus, if both the (1-0)S(0) and (1-0)S(1) lines are detected, a low $\eta$ value may be used as an indication of CR excitation in the cloud interior.

While ground-based slit-spectroscopy is not sufficiently sensitive to detect these lines (in clouds exposed to the mean FUV interstellar field, $I_{\rm UV} \approx 1$, where CR excitation is relatively important), an alternative observational strategy is to use a large-beam scanning Fabry-Perot filter.
This approach has the advantage that (a) the observing field of view is much larger, and thus the signal is gathered from a large fraction of the cloud area (\citealt{Luhman1994, Luhman1996}; see also the discussion in \citetalias{Bialy2020}, ``detectability" section).
Indeed, adopting this approach, \citet{Luhman1994} and \citet{Luhman1996} were able to detect very faint and extended emission of the (6-4)Q(1), (1-0)S(1), and (2-1)S(1) lines in various galactic PDRs.

A more robust determination 
of FUV and CR excitation
may be achieved by relying on a large number of H$_2$ transitions, including various ortho-H$_2$ and para-H$_2$ lines, and various vibrational states, $v=0,1,2$, etc. This approach has the advantage that it includes 
thee two CR excitation effects discussed above (i.e., (1) direct impact versus radiative cascade for CR vs FUV, and (2) different ratios due to the different temperatures in the PDR and inner cloud zone), plus, the fact that the analysis uses many independent lines, makes it less sensitive to observational errors and model uncertainties. 
Given an observed H$_2$ spectrum, fitting it with a thermo-chemical model that self-consistently calculates the thermal structure and the FUV and CR excitations (including both the exterior PDR zone and the inner CR-dominated region) will allow to reveal the contribution of CRs to the H$_2$ excitation, to robustly determine the values of $I_{\rm UV}$ and $\zeta$, and in turn to constrain the low-energy spectral slope of interstellar CR protons. 
As we demonstrated in this paper, for clouds illuminated by the typical interstellar radiation field, $I_{\rm UV} \approx 1$, this goal is not achievable from the ground, however, it is very feasible
 with future space observations, with upcoming JWST (\S \ref{sec: JWST}).
 In addition to the high SNR and the detection of many lines simultaneously, another advantage of  observations from space is that some lines are completely blocked by the atmosphere and can only be detected from space.
 Most notably is the (1-0)O(2) line which for CR excitation is predicted to be the brightest H$_2$ line (see Table 1 in \citetalias{Bialy2020}).

Complementing the NIR spectra with observations at shorter wavelengths may be very useful for constraining $I_{\rm UV}$. 
 For example, as discussed by \citet{Neufeld1996} H$_2$ lines in the red-visible are more readily absorbed by dust and thus preferentially trace the conditions in the cloud outer boundary layer (where FUV excitation dominates).
 Observations in the FUV, both of the H$_2$ FUV fluorescent lines as well as of the scattered continuum starlight, are useful for constraining the illuminating FUV radiation, as well as of other proprieties of the gas and dust in the PDR \citep[e.g.][see also the discussion in \S 4.2 in \citealt{Bialy2017b}]{Lee2006, Jo2017, Mattila2018}.

\end{appendix}
\end{document}